\documentstyle[twoside,fleqn,espcrc2,epsf]{article}

\newcommand{\be}{\begin{equation}}                                              
\newcommand{\ee}{\end{equation}}                                                
\newcommand{\half}{\frac{1}{2}}

\newcommand{\AmS}{{\protect\the\textfont2
  A\kern-.1667em\lower.5ex\hbox{M}\kern-.125emS}}

\title{
{\vspace{-3cm} \normalsize
\hfill \parbox{30mm}{DESY 94-222}}\\[25mm]
       The Electroweak Phase Transition:\\
       A non-perturbative Lattice Investigation
        \thanks{Combined talks given by Z.F. and K.J. at the international
                Symposium on Lattice Field Theory, Sept. 27.-Oct.1., 1994,
                Bielefeld.}}
\author{F. Csikor\address{Institute for Theoretical Physics,\\
        E\"otv\"os University, Budapest, Hungary},
        Z. Fodor
         \address{
        Deutsches Elektronen-Synchroton DESY, \\ 
        Notkestr. 85, 22603 Hamburg, Germany}\thanks{
         On leave from Inst. for Theor. Physics,
         E\"otv\"os University, Budapest, Hungary},
        \addtocounter{address}{-1}
        J. Hein\addressmark,
        \addtocounter{address}{-1} 
        K. Jansen\addressmark,
        \addtocounter{address}{-1}
        A. Jaster\addressmark,
        \addtocounter{address}{-1}
        I. Montvay \addressmark$^{,}$ 
        \address{
        CERN, Theoretical Physics Division, \\ 
        CH-1211 Geneva 23, Switzerland}}%

\begin{document}

\begin{abstract}
We present results obtained from a numerical investigation of
the electroweak phase transition in the SU(2)-Higgs model. The 
simulations are performed at two values of the Higgs boson mass,
$M_H\approx 20$ GeV and $M_H\approx 50$ GeV. While the phase transition
is of strongly first order at the smaller value of the Higgs mass it
weakens rapidly when the Higgs mass is increased. This is in qualitative
agreement with perturbation theory as the comparison of various
physical quantities shows.

\end{abstract}

\maketitle

\section{Introduction}
 The number of baryons is not conserved in                  
 the minimal standard model \cite{THOOFT}. Also CP is not an exact
 symmetry. If, in addition, the cosmological finite temperature {\em electroweak
 phase transition} would be of strong enough first order, all three
 of Sakharov's conditions for a generation of an asymmetry between baryons
 and antibaryons would be fulfilled.
 This offers the intriguing possibility that the observed baryon asymmetry 
 of the universe can be                   
 explained within the minimal standard model alone \cite{SHAPOSH}.           
 The resolution of this question is therefore a major challenge for             
 elementary particle physics.                                                   
                                                                                
 The standard calculational method for the study of the symmetry                
 restoring electroweak phase transition is resummed perturbation theory         
 \cite{CARRIN,BUFOHEWA,ARNESP,FODHEB}. While                               
 in the Higgs phase perturbation theory is expected to work well for            
 Higgs boson masses smaller than $M_H \sim 80$ GeV, 
 in the symmetric phase 
 irreparable infrared singularities            
 occur which prevent a quantitative control of graph resummation                
 \cite{LINDE}.                                                                  
 Indeed, the results of perturbation theory show bad convergence for
 physical quantities depending on characteristics of both phases.

 One non-perturbative approach
 for the study of the electroweak 
 phase transition, which we will adopt, 
 is based on numerical simulations.                           
 In the work described below, fermions and the U(1) gauge fields are
 omitted, which can be expected on general grounds to be a reasonable
 first approximation. One is therefore left with the SU(2)-Higgs model.                                                                                
 We will stay in the original four-dimensional theory without 
 dimensional reduction.             
 This has the advantage of keeping the number of bare parameters small          
 and not introducing any further approximations beyond the lattice              
 regularization. The results presented in these talks                                                                 
 have been published recently \cite{zoltankarl,CFHJJM,FHJJM}.            
 There a much more detailed discussion and a more extended
 list of references can be found.
 We restrict the present calculations to smaller Higgs                
 boson masses below 50 GeV.                                                     
 Since this region of parameters of the minimal standard model is               
 already excluded by experiments, our present scope is merely                   
 theoretical because we would like to check the validity of                     
 some other theoretical approximation schemes, e.g.~resummed                   
 perturbation theory.                                                           
 
 The lattice action of the SU(2) Higgs model is conventionally                  
 written as                                                                     

\[
S[U,\varphi] = \beta \sum_{pl}                                                  
\left( 1 - \frac{1}{2} {\rm Tr\,} U_{pl} \right)
\]
\[                                
 + \sum_x \left\{ \half{\rm Tr\,}(\varphi_x^+\varphi_x) +                        
\lambda \left[ \half{\rm Tr\,}(\varphi_x^+\varphi_x) - 1 \right]^2
 \right. 
\]
\be \label{action}
 -\left. \kappa\sum_{\mu=1}^4                                                            
{\rm Tr\,}(\varphi^+_{x+\hat{\mu}}U_{x\mu}\varphi_x)                            
\right\} \ .                                                                    
\ee
 Here $U_{x\mu}$ denotes the SU(2) gauge link variable, $U_{pl}$                
 is the product of four $U$'s around a plaquette and                            
 $\varphi_x$ is a complex $2 \otimes 2$ matrix in isospin space                 
 describing the Higgs scalar field and satisfying                               
$\varphi_x^+ = \tau_2\varphi_x^T\tau_2$.
 The bare parameters in the action are $\beta \equiv 4/g^2$ for                 
 the gauge coupling, $\lambda$ for the scalar quartic coupling and              
 $\kappa$ for the scalar hopping parameter related to the bare                  
 mass square $\mu_0^2$ by $\mu_0^2 = (1-2\lambda)\kappa^{-1} - 8$.              
 Throughout this paper we set the lattice spacing to one ($a=1$),               
 therefore all the masses and correlation lengths etc.\ will always be          
 given in lattice units, unless otherwise stated.                               
                                                                                
Since we are interested in the study of the symmetry restoring 
 phase transition as a function of temperature,                                 
 we use asymmetric lattices: the small temporal extensions                       
 $L_t=2,3,\ldots$ represent the discretized inverse temperature                 
 $L_t = 1/(aT)$.                                                                
 The other three (spatial) extensions of the lattice have to be much            
 larger, for reaching the thermodynamical limit.                                
 As stated before, we will stay at small Higgs boson masses.
 Hence for the bare quartic coupling we have chosen values near                 
 $\lambda = 0.0001$ (corresponding to $M_H\sim 18$ GeV) and 
 $\lambda = 0.0005$ (corresponding to $M_H\sim 50$ GeV).                                
 In the present paper the inverse temperature in lattice units will             
 be restricted to $L_t=2$ and $L_t=3$.                                          

\section{Monte Carlo simulation} 

 The simulations have been performed on the APE (Alenia Quadrics) computers           
 at DESY-IFH.                                                                       
It turned out that in the SU(2) Higgs model one encounters large
autocorrelation times already on small lattices of size $8^4$ if one
uses a standard Metropolis algorithm. We found that the 
expectation value of the Higgs
field length $\rho_x^2 = \frac{1}{2}\mbox{Tr}(\varphi_x^\dagger\varphi_x)$
shows the largest autocorrelation 
of all investigated quantities. By adopting an overrelaxation method
for the radial mode of the Higgs field \cite{zoltankarl} in combination with
standard overrelaxation and heatbath methods for the gauge fields and
the SU(2)-angle of the scalar field we were able to reach a substantial
reduction of the autocorrelation time as fig.~1 demonstrates.

\begin{figure}[htb]
\centerline{ \epsfysize=8.0cm \epsfxsize=8.5cm \epsfbox{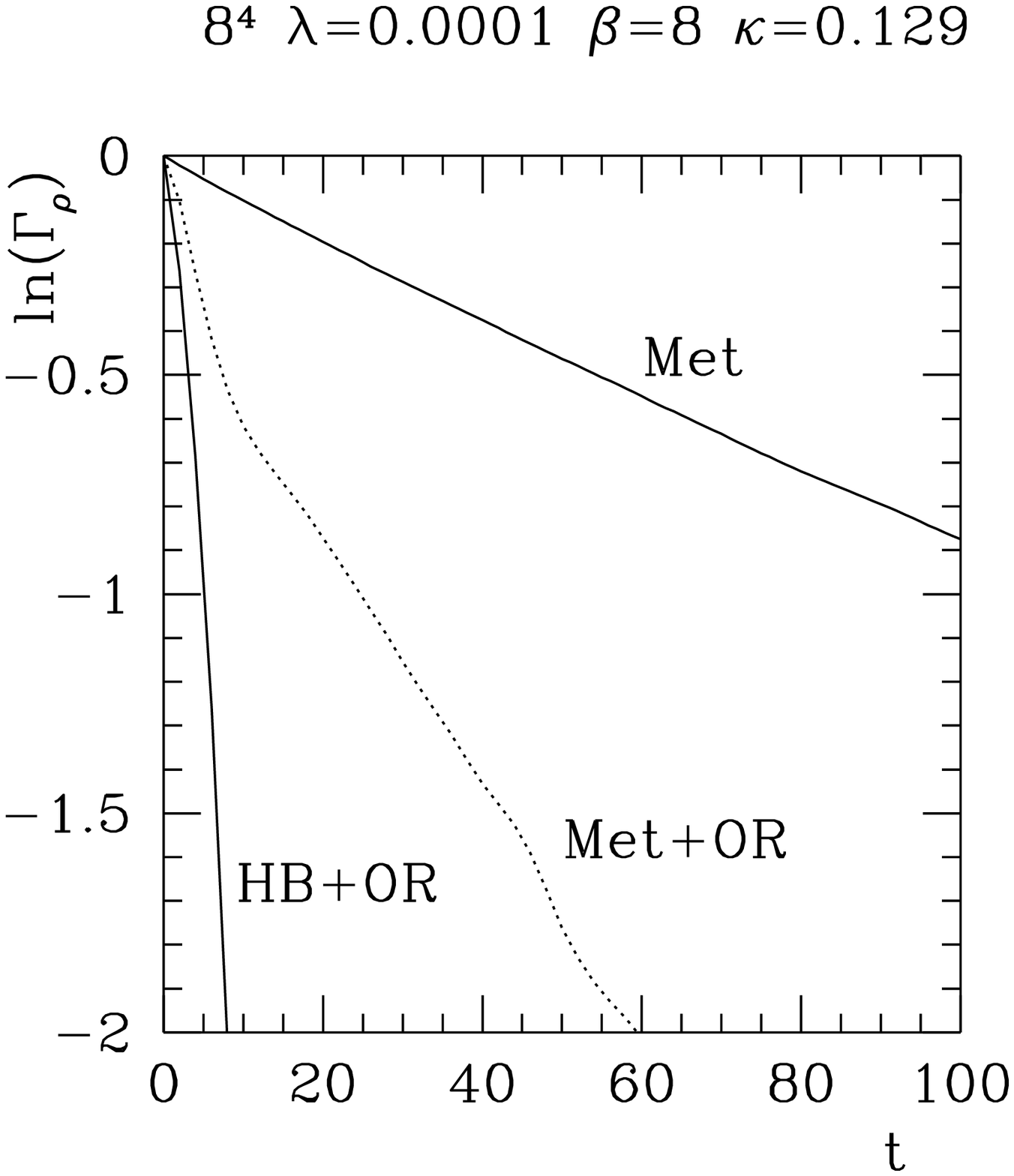}}
\vspace{-15mm}
\caption{The logarithm of the autocorrelation function for the length of the
scalar field $\rho$. Shown are different algorithms and their combinations, i.e.
Metropolis (Met), heatbath (HB) and overrelaxation (OR). Clearly a mixture
of HB and OR gives the smallest autocorrelation time.}
\vspace{-9mm}
\end{figure}

In \cite{FHJJM} we performed a search for an optimized mixing ratio 
of the various types of algorithms.                                 
This mixing ratio was then used for the simulations
described below.

\section{Phase transition points}

 A numerical simulation of the SU(2)-Higgs model should start by first          
 determining the phase transition points.                                       
 A change of $\kappa$ is reflected mainly in a change of the lattice            
 spacing $a$.                                                                   
 Therefore if one crosses the transition at fixed $\beta,\;\lambda$ by          
 changing $\kappa$, the essential change is in the physical temperature         
 $T = 1/(aL_t)$.                                                                
 (The physical volume is assumed to be large enough such that its change        
 with $a^3$ is not important.)                                                  
 Thus we are looking for the phase transition in the hopping parameter          
 at $\kappa=\kappa_c$, for fixed $\beta,\;\lambda$.                             
                                                     
 We found that for searching the transition point the gauge invariant           
 effective potential 
depending on the square of the Higgs field length $\overline{\rho}^2$
can be very helpful. 
The gauge invariant effective potential to 1-loop order in the broken
phase is given by

\be \label{gaugeinvarianteffpot}                                                               
  V_{1-loop} =  V_{tree} 
\ee
\[
 + \int_{-\pi}^{\pi}\frac{d^4 k}{(2\pi )^4}\left\{                               
  \frac{9}{2} \ln (\hat{k}^2 + m_g^2)                                           
+ \frac{1}{2} \ln (\hat{k}^2 + m_\phi^2 ) \right\} \ , 
\]
%
 where the masses are related to the parameters in the lattice action           
 (\ref{action}) and $\bar{\rho}^2$ by                                            
\be  
m_g^2   =  \frac{1}{2}\kappa g^2 \bar{\rho}^2  \  ,  \hspace{3em}
m^2_\phi  =  \frac{4}{\kappa}\lambda \bar{\rho}^2                             
\ee
 and the momenta in the lattice integrals are                                   
 $\hat{k}^2 = \sum_\mu[2-2\cos (k_\mu )]$.               
The tree level potential reads
\be 
V_{tree}(\bar{\rho}^2) = (1-8\kappa) \bar{\rho}^2                               
+\lambda (\bar{\rho}^2-1)^2.                                                    
\ee
 For the computation of the gauge invariant effective potential                 
 on a finite lattice of size $L_x \cdot L_y \cdot L_z \cdot L_t$ the            
 integrals in (\ref{gaugeinvarianteffpot}) 
have been replaced by the corresponding             
 lattice sums.                                                                  
 Following the experience in QCD, we also used the                  
 mean field improved gauge coupling $g\rightarrow g/\sqrt{U_{pl}}$,             
 with $U_{pl}$ the measured plaquette value.                                    
 This choice for the gauge coupling  appeared to be very useful to              
 achieve better agreement with simulation results. 


An alternative method for the determination of the phase transition points
is the so called two-coupling method. 
Here the lattice in the $z$-direction is divided into two halves, each of
which is equipped with a different hopping parameter $\kappa$ such
that one of them corresponds to the symmetric and the other to the
broken phase.
If the $z$-direction of the lattice
is long enough to accomodate a pair of interfaces and the configuration 
stays for many autocorrelation times in the mixed state the free energies
are such that the situation is stable against transitions to a unique 
phase.  Thus the parameter sets of the two phases give a lower and an upper            
 bound on the transition parameters. The critical hopping parameter is then
determined by the average of the best lower and upper $\kappa$-value.
The values of $\kappa_c$ computed in this way are in very good agreement with
the one obtained from the gauge invariant potential at $\lambda = 0.0001$.
For $\lambda = 0.0005$ we observe a $L_T$-independent shift by
$\Delta\kappa_c = 0.00025(1)$ (see table~2 in \cite{FHJJM}).

By far the most accurate determination of the critical value of $\kappa$ is 
obtained from the equal height criterion of the canonical distribution of the
action density. The distributions have been obtained by a combination
of the multicanonical method with $\kappa$-reweighting. We show the 
distribution at $\kappa_c = 0.128307$ in fig.~2. Note that the equal
height criterion is roughly equal to the equal area condition in this case.

\begin{figure}[htb]
\vspace{-10mm}
\centerline{ \epsfysize=10.0cm \epsfxsize=8.0cm \epsfbox{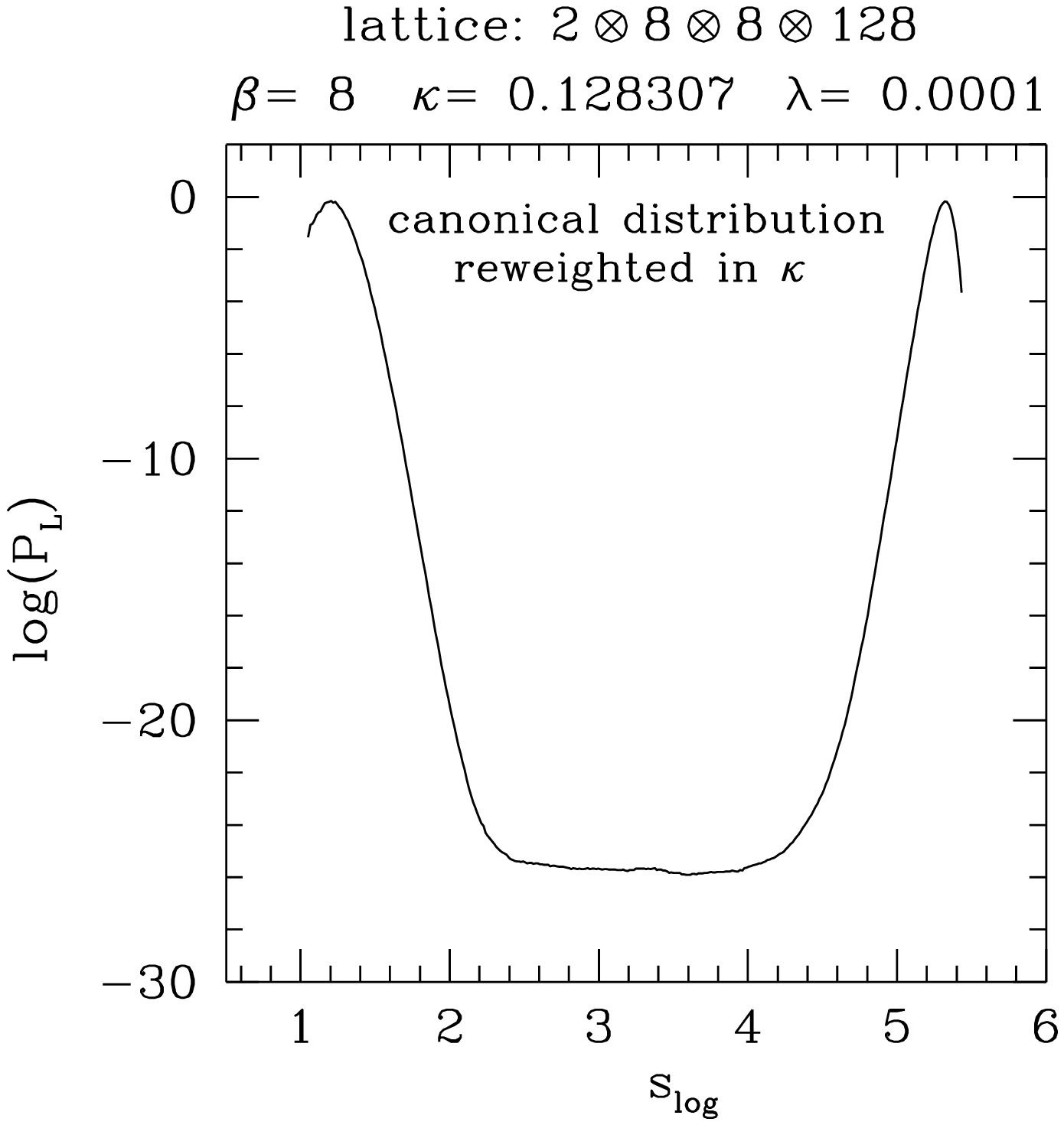}}
\vspace{-35mm}
\caption{Distribution of the action density.}
\vspace{-7mm}
\end{figure}


\section{Masses}
                                              
 The physical Higgs mass $M_H$ can be extracted from correlators                
 of quantities as the site variable                                             
\be \label{higgsmass}                                                               
R_x \equiv \half{\rm Tr\,}(\varphi_x^+\varphi_x)                                
\equiv \rho_x^2 \ .                                                             
\ee                                                                             
There are alternative choices which have also been used.                                                                              
 The W-boson mass $M_W$ can be obtained similarly from the composite            
 link fields ($r,k=1,2,3$)                                                      
\be \label{wmass}                                                               
W_{xrk} \equiv                                                                  
\half {\rm Tr\,}(\tau_r \alpha^+_{x+\hat{k}}U_{xk}\alpha_x) \ .                 
\ee                                                                             
 Due to lattice symmetry only the diagonal correlators in                       
 $(r,k)$ are non-zero and they can be averaged before extracting                
 masses.                                                                        
                                                                                
 Masses were extracted from the connected correlators by least square           
 fits by a single $\cosh$ or $\cosh$ + $\rm constant$.                              
 Constant contributions are possible in the Higgs channels but were             
 most of the time negligible, because they are of the order                     
 $\exp(-M_H L_t)$ and our lattices usually have large enough time               
 extension $L_t$.                                                               
                                                                                
 Statistical errors on masses were always estimated by subdividing              
 the data sample into subsamples.                                               
 Performing the fits in subsamples gives estimates of standard                  
 deviations of fit parameters.                                                  
 This is particularly straightforward on the Quadrics Q16,                      
 because these lattices can be simulated on 8 nodes                             
 which is repeated 16 times with independent sequences of random                
 numbers.                                                                       
 The 16 parallel sets of statistically independent results can be used          
 for the estimate of fit parameter errors.                                      

 An important effect of finite temperatures is the change in                    
 correlation lengths.                                                           
 This is particularly interesting at a first order phase transition,            
 where correlation lengths stay finite and may eventually differ                
 in the two metastable states.                                                  
                                                                                
 In our numerical simulations we make use of the fact that on a large           
 enough lattice the metastability at the first order phase transition           
 becomes so strong that in a finite amount of computer time the system          
 stays in the phase it started from.                                            
Therefore, the masses can be extracted in the symmetric and the broken
phase separately.

 There is a clear signal for a discontinuity of $m_W$ and $m_H$                 
 between the two sides of the phase transition
 which shrinks with increasing (zero temperature) Higgs
 mass as shown in fig.~3 for the case of $m_H$. 
 Thus we have a first indication that the
 phase transition becomes weaker for larger Higgs masses.
 The values of $m_{W,H}$ are rather small, therefore the lattice                
 volumes may be too small and
 the infinite volume values may still be somewhat different.              
\begin{figure}[htb]
\centerline{ \epsfysize=7.0cm \epsfxsize=7.5cm \epsfbox{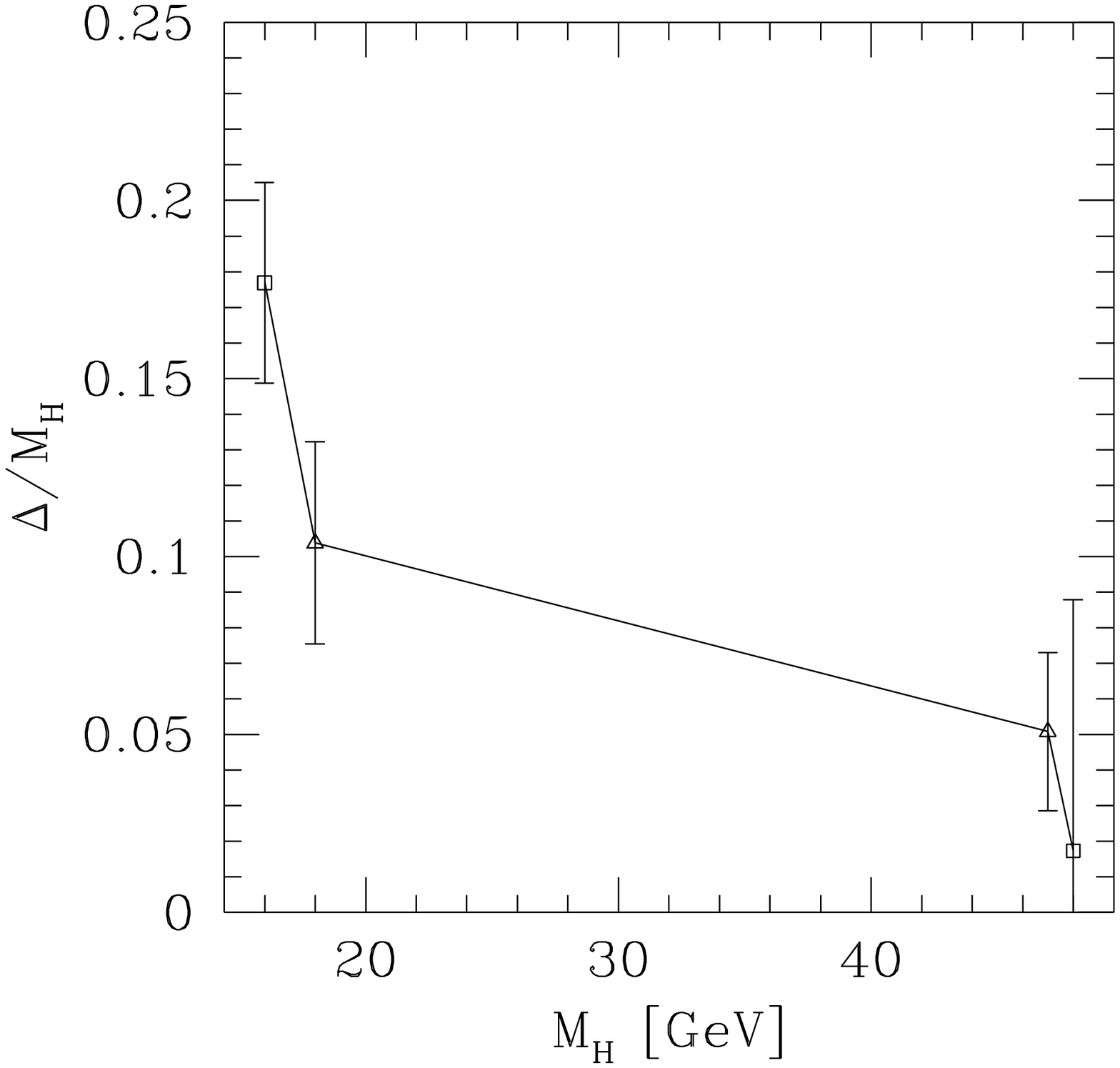}}
\vspace{-10mm}
\caption{Finite temperature mass gap $\Delta=|m_{sym}-m_{br}|$
with $m_{sym}$ and $m_{br}$ the finite temperature Higgs masses 
in the symmetric and
the broken phase, respectively. Triangles correspomd to $L_t =2$ and squares
to $L_t=3$.}
\vspace{-7mm}
\end{figure}

\section{Renormalization group trajectories}  

 The continuum limit of quantum field theories on the lattice is taken          
 along renormalization group trajectories, also called {\em lines of            
 constant physics} (LCP's).                                                     
 Going to the continuum limit along such lines in bare parameter space,         
 the same physical theory is reproduced with increasing precision.              
 In the SU(2) Higgs model, in order to define the LCP's, one has to             
 keep fixed the values of two independent renormalized couplings,               
 the renormalized gauge ($g_R^2$) and the renormalized 
 quartic ($\lambda_R=M_W^2/(32M_H^2g_R^2)$) couplings.                                                       
 The parameter characterizing the points along LCP's can be               
 chosen as                                                                      
$\tau \equiv \log M_W^{-1}$.                                                 
 (Remember that in this paper the lattice spacing is set to $a=1$,              
 therefore $M_W$ is measured here in lattice units.)                            
                                                                                
 Since our numerical simulations are performed in the weak coupling             
 region of parameter space, the change of the bare couplings                    
 $g^2 \equiv 4/\beta$ and $\lambda_0 = \lambda/4\kappa^2$ 
 along LCP's should be well approximated by the solutions of the                   
 one-loop perturbative renormalization group equations:                         
\[
\frac{dg^2(\tau)}{d\tau} = -\frac{43}{48\pi^2}g^4                          
 + {\cal O}(\lambda_0^3,\lambda_0^2g^2,\lambda_0g^4,g^6)  \ ,                                                                     
\]
\begin{eqnarray} \label{lcp}
\frac{d\lambda_0(\tau)}{d\tau} & = &\frac{1}{16\pi^2} \left[                       
96\lambda_0^2 + \frac{9}{32}g^4 - 9\lambda_0 g^2 \right] \nonumber\\
 & + & {\cal O}(\lambda_0^3,\lambda_0^2g^2,\lambda_0g^4,g^6) \ .               
\end{eqnarray}
 The integration of these equations can be started at the transition            
 points of the $L_t=2$ lattices.                                                
 At a distance $\Delta\tau=\log(L_t/2)$ ($L_t=3,4,5,\ldots$) in                 
 parameter $\tau$ one obtains the corresponding values of                       
 $(g^2,\lambda_0)$ for the transition points of lattices with temporal          
 extension $L_t$. To give some impression of how the bare parameters change,
going from $L_t=2$ to $L_t=3$ changes $\lambda=0.0001$ to $\lambda=0.00011$
and $\beta=8.0$ to $\beta=8.147$, respectively.                                                               
                                                                                
 In order to have a perturbative prediction also for the change of              
 the third bare parameter $\kappa$, one can make use of the one-loop            
 perturbative invariant effective potential.                                    
 The measured values of                  
 $\kappa_c$ are very well reproduced by the one-loop invariant                  
 effective potential for $\lambda=0.0001$ and show 
 a parallel shift between             
 prediction and measurements: the prediction is too low by                      
 $\Delta\kappa_c = 0.00025(1)$ at $\lambda=0.0005$.                        
 As a consequence, the derivatives of $\kappa$ along the LCP's                
 can be well determined.                                                        
 The estimates of the LCP's by 
 one-loop perturbation theory work amazingly well.            
 We determined the renormalized gauge coupling as usual from the
 static potential computed from Wilson loops measured at zero
 temperature. At $\lambda =0.0001$ 
 we find for $L_t=2$ and $L_t=3$ the values of $g_R^2$ to be $0.564(6)$ and
 $0.5597(10)$, respectively. At $\lambda=0.0005$ we obtained
$g_R^2(L_t=2) = 0.5763(10)$ and 
$g_R^2(L_t=3) = 0.5651(13)$. Thus we see that the renormalization is first
of all quite small and secondly that scaling violations are tiny thus confirming
the prediction of the 1-loop renormalization group equations. A similar picture
is obtained from the ratio $R_{HW}$ of the zero temperature Higgs and W-masses:
$R_{HW}(L_t=2) = 0.222(12)$, $R_{HW}(L_t=3) = 0.201(5)$ for $\lambda=0.0001$ and
$R_{HW}(L_t=2) = 0.593(19)$, $R_{HW}(L_t=3) = 0.606(24)$ for $\lambda=0.0005$.

                                                                                
\section{Latent heat}                    
 An important characteristic feature of first order phase transitions           
 is the latent heat, i.e.\ the discontinuity of the energy density              
 $\epsilon$.                                                                    
 As it has been shown in \cite{CFHJJM}, the latent heat                         
 $\Delta\epsilon$ in the SU(2) Higgs model is given by                          
\begin{eqnarray} \label{latentheat}                                                                
\frac{\Delta\epsilon}{T_c^4} & = & L_t^4                                            
\left\langle \frac{\partial\kappa}{\partial\tau}                                
\cdot 8\Delta L_{\varphi,x\mu} \right. \nonumber \\                                                
& - & \left. \frac{\partial\lambda}{\partial\tau} \cdot \Delta Q_x                         
- \frac{\partial\beta}{\partial\tau}                                            
\cdot 6\Delta P_{pl} \right\rangle \ .                                          
\end{eqnarray}                                                                             
 Here we used the gauge invariant variables 
\[
L_{\varphi,x\mu} \equiv                                                         
\half {\rm Tr\,}(\varphi^+_{x+\hat{\mu}}U_{x\mu}\varphi_x) \; ,
\]
\be \label{jumpquantities} 
Q_x \equiv (\rho_x^2 - 1)^2 \;\; \mbox{and}\;\;                                                 
P_{pl} \equiv 1 - \half {\rm Tr\,} U_{pl} \ .                                   
\ee                                                                             

The jumps of the above quantities can be obtained 
 by performing the simulations on large enough            
 lattices, where the strong metastability of phases can be exploited.           
 For $\Delta\epsilon$ in eq.~(\ref{latentheat}) 
 besides the discontinuities of global quantities                          
the derivatives of bare                 
 parameters along the LCP's are needed.                                         
 These have been determined in the way discussed in the previous section.                             
 We want to note that the main contribution to the error of the
latent heat comes from the error
 of $\partial\kappa/\partial\tau$ as evaluated           
 from the one-loop invariant effective potential and from numerical             
 simulation data. The results for the latent heat read for $\lambda=0.0001$

\[                                                                              
\frac{\Delta\epsilon}{T_c^4}  = 1.81(29)\;\;   \ , \;               
\frac{\Delta\epsilon}{T_c^4}  = 1.57(13)   \                                                                     
\]
and for $\lambda=0.0005$,
\[                                                                              
\frac{\Delta\epsilon}{T_c^4}  = 0.132(17)\;\; \ ,  \;                                                         
\frac{\Delta\epsilon}{T_c^4}  = 0.122(9)   \ .               
\]                                                                             
Here the first numbers correspond to $L_t=2$ and the second
numbers to $L_t=3$.
 We see here that within errors the results on $L_t=2$                     
 and $L_t=3$ lattices coincide, again pointing in the direction that
scaling violations are small.                                                 
                                                                                
                                                                                
\section{Interface tension} 
                                                                                
 At the transition point of the electroweak phase transition                    
 mixed states can appear, where different bulk phases are separated by          
 interfaces.                                                                    
 The interface tension, $\sigma$, is the free energy per unit area
 associated         
 with these interfaces.                                                         
 The dynamics of a first order phase transition is to a large extent            
 determined by the latent heat and the interface tension.                   
 In this subsection we present our results for the interface                    
 tension obtained by the histogram method                          
 for $\lambda=0.0001$ and $L_t=2$.                                          
                                                                                
 At $\kappa_c$ the probability distribution of an order parameter               
 (e.g.~action density $s \equiv S/\Omega$ defined by               
 (\ref{action}), or link variable $L_\varphi$ in (\ref{jumpquantities})) 
develops two peaks.                                                       
 They correspond to pure phases and the suppressed configurations               
 between the peaks are dominated by mixed states, where the phases              
 are separated by interfaces.                                                   
 Defining $\kappa_c$ by the equal height signal, the suppression                
 at infinite volume is given by the interface tension                           
\be \label{simgainfty}                                                                            
\sigma_\infty=                                                                  
\lim_{\Omega \rightarrow \infty} \sigma_\Omega \;\; ,                              
\sigma_\Omega={1 \over 2 L_x L_y L_t} \log{p_{max} \over p_{min}} \ ,           
\ee                                                                             
 where $p_{max}$ corresponds to the heights of the peaks and                    
 $p_{min}$ to the minimum in between.                                           
 In the case of elongated lattices, as was used here,
 the finite size corrections can be given explicitly
\be \label{finitesize}                                                               
\sigma_\infty=\sigma_\Omega+{1 \over L_{xy}^2 L_t}                              
(c+{3 \over 4} \log L_z -{1 \over 2} \log L_{xy}) \ ,                           
\ee                                                                             
 where $L_x=L_y \equiv L_{xy}$, $L_z$ is the longest extension and $c$          
 is an unknown constant.                                                        
 The obvious advantage of this choice is that practically all mixed             
 configurations contain two surfaces perpendicular to the $z$                   
 direction. As is demonstrated in fig.~2, the distribution of the action
 density can be well measured, using the multicanonical simulation
technique. As a result, we find                                             
 the finite volume interface tensions to be                            
 $\sigma_\Omega/T_c^3=0.63(3)$ for $\Omega=2\cdot 4^2 \cdot  128$ and           
 $\sigma_\Omega/T_c^3=0.80(2)$ for $\Omega=2\cdot 8^2 \cdot  128$,              
 respectively.                                                                  
 Combining these two values one gets from (\ref{finitesize})               
\be \label{sigmaresulte}                                                  
\left( \frac{\sigma_\infty}{T_c^3} \right) = 0.83(4) \ .          
\ee                                                                             
 This result is                                  
 in very good agreement with the results of the                                 
 two-$\kappa$ method discussed in \cite{FHJJM}.

\vspace{-2mm}
\section{Conclusion} 

The non-perturbative lattice computations discussed in this paper
demonstrate the first order nature of the electroweak
phase transition for Higgs boson masses $M_H < 50$ GeV.
 The strongest indication for this is the two-peak structure of                 
 order parameter distributions (see e.g.~fig.~2) showing
 a pronounced two peak structure. The lattice data follow the
trend that the phase transition weakens substantially when the Higgs mass
is increased. 
This can be seen in quantities characterizing the phase transition like the
latent heat or the surface tension. 
The rate of decrease is qualitatively the same as given by two-loop            
resummed perturbation theory \cite{FODHEB,CFHJJM,FHJJM}.

Another interesting observation is the smallness of the scaling violations.
These can be estimated by studying physical quantities as a function of the
temporal extension of the lattice. We find that, within the error bars,
the renormalized couplings $\lambda_R$ and $g_R^2$ stay constant as predicted
by the 1-loop renormalization group equation. In addition, dimensionless
quantities like $\Delta\epsilon/T_c^4$ or $T_c/M_H$ 
(see \cite{FHJJM} for the latter quantity) remain almost constant.
 The apparent general tendency of small scaling violations already at           
 such small temporal lattice extensions is consistent with the                  
 expected dominance of the lowest Matsubara modes motivating the                
 dimensional reduction.                                    


\end{document}